\documentstyle[12pt]{article}
\input epsf
\begin{document}
\title{ 
\hfill hep-ph/9907390\\ \vskip .5truecm
\hfill PRL-TH-99/004 \\ \vskip 1.0truecm
{\large {\bf Neutrino Anomalies in an Extended Zee Model}}} 
\vskip 2.0truecm
\author{ Anjan S. Joshipura and Saurabh D. Rindani\\
{\ns\it Theoretical Physics Group, Physical Research Laboratory,}\\
{\ns\it Navarangpura, Ahmedabad, 380 009, India.}}
\date{}
\def\ap#1#2#3{           {\it Ann. Phys. (NY) }{\bf #1} (19#2) #3}
\def\arnps#1#2#3{        {\it Ann. Rev. Nucl. Part. Sci. }{\bf #1} (19#2) #3}
\def\cnpp#1#2#3{        {\it Comm. Nucl. Part. Phys. }{\bf #1} (19#2) #3}
\def\apj#1#2#3{          {\it Astrophys. J. }{\bf #1} (19#2) #3}
\def\asr#1#2#3{          {\it Astrophys. Space Rev. }{\bf #1} (19#2) #3}
\def\ass#1#2#3{          {\it Astrophys. Space Sci. }{\bf #1} (19#2) #3}

\def\apjl#1#2#3{         {\it Astrophys. J. Lett. }{\bf #1} (19#2) #3}
\def\ass#1#2#3{          {\it Astrophys. Space Sci. }{\bf #1} (19#2) #3}
\def\jel#1#2#3{         {\it Journal Europhys. Lett. }{\bf #1} (19#2) #3}

\def\ib#1#2#3{           {\it ibid. }{\bf #1} (19#2) #3}
\def\nat#1#2#3{          {\it Nature }{\bf #1} (19#2) #3}
\def\nps#1#2#3{          {\it Nucl. Phys. B (Proc. Suppl.) }
                         {\bf #1} (19#2) #3} 
\def\np#1#2#3{           {\it Nucl. Phys. }{\bf #1} (19#2) #3}
\def\pl#1#2#3{           {\it Phys. Lett. }{\bf #1} (19#2) #3}
\def\pr#1#2#3{           {\it Phys. Rev. }{\bf #1} (19#2) #3}
\def\prep#1#2#3{         {\it Phys. Rep. }{\bf #1} (19#2) #3}
\def\prl#1#2#3{          {\it Phys. Rev. Lett. }{\bf #1} (19#2) #3}
\def\pw#1#2#3{          {\it Particle World }{\bf #1} (19#2) #3}
\def\ptp#1#2#3{          {\it Prog. Theor. Phys. }{\bf #1} (19#2) #3}
\def\jppnp#1#2#3{         {\it J. Prog. Part. Nucl. Phys. }{\bf #1} (19#2) #3}

\def\rpp#1#2#3{         {\it Rep. on Prog. in Phys. }{\bf #1} (19#2) #3}
\def\ptps#1#2#3{         {\it Prog. Theor. Phys. Suppl. }{\bf #1} (19#2) #3}
\def\rmp#1#2#3{          {\it Rev. Mod. Phys. }{\bf #1} (19#2) #3}
\def\zp#1#2#3{           {\it Zeit. fur Physik }{\bf #1} (19#2) #3}
\def\fp#1#2#3{           {\it Fortschr. Phys. }{\bf #1} (19#2) #3}
\def\Zp#1#2#3{           {\it Z. Physik }{\bf #1} (19#2) #3}
\def\Sci#1#2#3{          {\it Science }{\bf #1} (19#2) #3}
\def\n.c.#1#2#3{         {\it Nuovo Cim. }{\bf #1} (19#2) #3}
\def\r.n.c.#1#2#3{       {\it Riv. del Nuovo Cim. }{\bf #1} (19#2) #3}
\def\sjnp#1#2#3{         {\it Sov. J. Nucl. Phys. }{\bf #1} (19#2) #3}
\def\yf#1#2#3{           {\it Yad. Fiz. }{\bf #1} (19#2) #3}
\def\zetf#1#2#3{         {\it Z. Eksp. Teor. Fiz. }{\bf #1} (19#2) #3}
\def\zetfpr#1#2#3{         {\it Z. Eksp. Teor. Fiz. Pisma. Red. }{\bf #1} (19#2) #3}
\def\jetp#1#2#3{         {\it JETP }{\bf #1} (19#2) #3}
\def\mpl#1#2#3{          {\it Mod. Phys. Lett. }{\bf #1} (19#2) #3}
\def\ufn#1#2#3{          {\it Usp. Fiz. Naut. }{\bf #1} (19#2) #3}
\def\sp#1#2#3{           {\it Sov. Phys.-Usp.}{\bf #1} (19#2) #3}
\def\ppnp#1#2#3{           {\it Prog. Part. Nucl. Phys. }{\bf #1} (19#2) #3}
\def\cnpp#1#2#3{           {\it Comm. Nucl. Part. Phys. }{\bf #1} (19#2) #3}
\def\ijmp#1#2#3{           {\it Int. J. Mod. Phys. }{\bf #1} (19#2) #3}
\def\ic#1#2#3{           {\it Investigaci\'on y Ciencia }{\bf #1} (19#2) #3}
\def\tp{these proceedings}
\def\pc{private communication}
\def\ip{in preparation}
\relax

\newcommand{\GeV}{\,{\rm GeV}}
\newcommand{\MeV}{\,{\rm MeV}}
\newcommand{\keV}{\,{\rm keV}}
\newcommand{\eV}{\,{\rm eV}}
\newcommand{\Tr}{{\rm Tr}\!}
\renewcommand{\arraystretch}{1.2}
\newcommand{\beq}{\begin{equation}}
\newcommand{\eeq}{\end{equation}}
\newcommand{\beqa}{\begin{eqnarray}}
\newcommand{\eeqa}{\end{eqnarray}}
\newcommand{\ba}{\begin{array}}
\newcommand{\ea}{\end{array}}
\newcommand{\bmat}{\left(\ba}
\newcommand{\emat}{\ea\right)}
\newcommand{\refs}[1]{(\ref{#1})}
\newcommand{\ler}{\stackrel{\scriptstyle <}{\scriptstyle\sim}}
\newcommand{\ger}{\stackrel{\scriptstyle >}{\scriptstyle\sim}}
\newcommand{\lag}{\langle}
\newcommand{\rag}{\rangle}
\newcommand{\ns}{\normalsize}
\newcommand{\cm}{{\cal M}}
\newcommand{\gr}{m_{3/2}}
\newcommand{\p}{\partial}

\def\rp{ $R_P$} 
\def\321{$SU(3)\times SU(2)\times U(1)$}
\def\tl{{\tilde{l}}}
\def\tL{{\tilde{L}}}
\def\bd{{\overline{d}}}
\def\tL{{\tilde{L}}}
\def\a{\alpha}
\def\b{\beta}
\def\g{\gamma}
\def\c{\chi}
\def\d{\delta}
\def\D{\Delta}
\def\db{{\overline{\delta}}}
\def\Db{{\overline{\Delta}}}
\def\e{\epsilon}
\def\l{\lambda}
\def\n{\nu}
\def\m{\mu}
\def\nt{{\nu_{\tau}}}
\def\p{\phi}
\def\P{\Phi}
\def\x{\xi}
\def\r{\rho}
\def\s{\sigma}
\def\t{\tau}
\def\th{\theta}
\def\ne{\nu_e}
\def\nm{\nu_{\mu}}
\def\rp{$R_P$}
\def\mp{$M_P$}
\def\mgut{M_{GUT}}
\def\emt{$L_e-L_\m-L_{\tau}$ }     
\renewcommand{\Huge}{\Large}
\renewcommand{\LARGE}{\Large}
\renewcommand{\Large}{\large}
\maketitle
\vskip 2.0truecm
\begin{abstract}
We discuss an extended $SU(2)\times U(1)$ model which naturally leads to
mass scales and mixing angles relevant for understanding both the solar and
atmospheric neutrino anomalies. No right-handed neutrinos are introduced in the
model. 
The model uses a softly broken \emt symmetry.
Neutrino masses arise only at the loop level. The one-loop
neutrino masses which arise as in the Zee model 
solve the atmospheric neutrino anomaly while breaking of
\emt generates at two-loop order a mass splitting
needed for the vacuum solution of the solar neutrino
problem. A somewhat different model is possible which 
accommodates the large-angle MSW resolution of the solar neutrino problem.
\end{abstract}

Recent results on atmospheric neutrinos from SuperKamiokande
\cite{sk} seem to supply definite evidence that muon neutrinos oscillate into
either tau neutrinos or some species of sterile neutrinos. The evidence that 
solar neutrinos ($\nu_e$) oscillate into some other species has also been
mounting. Simultaneous understanding of these experiments needs two 
hierarchical mass scales  $\D_A$ and $ \D_S$
in the neutrino sector \cite{smir1}. The standard picture with only
three
generations
of light neutrinos can accommodate two mass scales. But theoretical
understanding of the pattern of
masses and mixings within conventional pictures of neutrino mass generation 
is not
straightforward.
Basically, apart from understanding why neutrino masses are small one
would also need to have theoretical understanding of the ({\em i})
two hierarchical mass scales and ({\em ii}) presence of the one or two
large mixing angles in the leptonic Kobayashi-Maskawa matrix. The
requirements ({\em i}) and ({\em ii}) becomes particularly stringent in
case of the vacuum oscillations (VO) solution \cite{vac} for the solar
neutrino
problem which needs   ${\Delta_S\over\Delta_A}\sim
10^{-8}$ and large mixing angle\footnote{Some of the models or textures
leading to vacuum solution
for the solar neutrino anomalies are presented in \cite{barb,just}.
More detailed list of references can be found in
\cite{smir1}.}.

One of the  possibilities for understanding the smallness of the neutrino
mass is to assume that it arises radiatively\footnote{A review and references
for models with radiatively generated
neutrino masses can be found in
\cite{mb}. See also \cite{baren}.}. An attractive
possibility is that $\D_A$ is
small because it arises at one loop in perturbation theory and $\D_S$ is
still smaller because it arises at the two-loop level. We describe here
precisely such a scheme in the context of an $SU(2)_L \times U(1)$ electroweak
gauge theory. 
 
The Zee model \cite{zee} provides
a particularly nice example of radiative neutrino mass generation. It
extends only the Higgs sector of the standard model (SM)  in a way
that admits lepton number
violation but no neutrino masses at the tree level. In the
most general situation, all
three neutrinos obtain their masses at the one-loop level
in this model. But the predicted spectrum and the mixing pattern is quite
constrained. In spite of that, the Zee model has been argued 
recently \cite{glashow} to provide a very good zeroth order approximation
to the required
neutrino spectrum for definite choice of parameters of the model.
Specifically, it was found \cite{glashow} that the only way in which one could
understand the hierarchical mass scales and the required 
mixing pattern in this model is to augment it with a (global)
$L_e-L_\m-L_\t$
symmetry \footnote{Another possible interpretation of the Zee
model is that it admits the dark matter and the atmospheric mass scales
\cite{smir2}.
In this case it can explain the $\nu_\m-\nu_\t$ and possible  $\nu_\m-\nu_e$ 
oscillations seen at SuperKamiokande and LSND respectively. But it cannot
accommodate a solution to the solar neutrino
problem unless a singlet neutrino is invoked \cite{smir2,gaur}.}.
This symmetry has been earlier recognized \cite{barb}
to lead to one and possibly two large mixing angles and two  degenerate
and one massless neutrinos. The muon neutrino deficit can be explained in
the model by  identifying  the degenerate mass 
with the atmospheric scale.
Moreover, the  bimaximal mixing
pattern \cite{bim} possible in this case 
provides an understanding of the absence of the electron neutrino
oscillation
at the atmospheric scale found at SuperKamiokande and at CHOOZ. The smallness
of the atmospheric scale also becomes understandable in the Zee model due to
its radiative origin. 

While providing an understanding of all the features of the atmospheric
neutrino results,
the Zee model plus $L_e-L_\m-L_\t$ symmetry fails to lead to the required
solar scale. This needs a small breaking of the $L_e-L_\m-L_\t$ symmetry.
Our aim here is to discuss a specific
extension of the Zee model in which the hierarchical ${\Delta_S\over
\Delta_A}$ gets generated at the two-loop level. The value of the 
${\Delta_S\over\Delta_A}$ obtained in the model would lead to vacuum
oscillations  as solution to the solar neutrino scale naturally but
it is also possible to obtain the large-angle MSW (LAMSW)
\cite{lamsw} solution in this
type of picture. 

The model described below contains a singly charged Higgs as in the
original model of Zee \cite{zee} and a doubly charged Higgs as in
model by Zee \cite{zee2} and by Babu \cite{babu}. The added feature here
is 
that the Yukawa
couplings conserve total lepton number $L$, as well as $L' \equiv
L_e-L_\m-L_\t$.\footnote{Similar models have been discussed \cite{rad} in
the past in the
context of a 17 keV neutrino or hot dark matter. All these models
invoked a  sterile neutrino in addition to the three light
neutrinos.}

The model is based on the gauge group $SU(2)_L\times U(1)\times SU(3)_C$, and
has the same fermion content as the standard model (SM). Thus, no singlet 
right-handed neutrinos are introduced. We introduce, in addition to the SM
$SU(2)_L$ doublet scalar field $\phi_1$, another doublet $\phi_2$, and
$SU(2)_L$
singlets $h^+$ and $k^{++}$ which carry respectively charges $+1$ and $+2$. In
addition to the total lepton number $L\equiv L_e+L_\m+L_\t$, an additional
global symmetry corresponding to $L'\equiv L_e-L_\m-L_\t$ is introduced. The
quantum numbers carried by scalar fields are shown in
Table 1.

\begin{table}
\begin{center}
\begin{tabular}{|c|c|c|c|c|}
\hline
&$SU(2)_L$&$Y$&$L$&$L'$\\
\hline
$\phi_{1,2}$&\underline{2}&$-\frac{1}{2}$&0&0\\
$h^+$&\underline{1}&$+1$&$-2$&0\\
$k^{++}$&\underline{1}&$+2$&$-2$&$-2$\\
\hline
\end{tabular}
\caption{Quantum numbers of the scalar fields under the various groups}
\end{center}
\end{table}

The Yukawa couplings of the leptons consistent with all symmetries are
\beqa
-{\cal L}_Y & =& g^{(a)}_{ij} \overline{L_{iL}}\phi_a e_{jR}
 + f_{e\mu}
\left(\overline{e^c_R}\nu_{\m L} - \overline{\m^c_R}\n_{eL}\right)h^+ \nonumber
\\ && 
+ f_{e\t}
\left(\overline{e^c_R}\nu_{\t L} - \overline{\t^c_R}\n_{eL}\right)h^+
+ h_{ee}\overline{e^c_R}e_Lk^{++} + {\rm H.c.},
\eeqa
where $g^{(a)}_{ij}$ ($a=1,2\; ; i,j=1,2,3$) have vanishing (12), (13),
(21) and (31)
elements. Vacuum
expectation values of $\phi_a$ generate the charged-lepton mass matrix, which
gives masses to all charged leptons. While mixing between the second and third
generations is permitted in this sector, there is no mixing between the
first and second or
third generations at tree level. The neutrinos are massless at tree
level.

Apart from the usual bilinear and quartic terms in
the fields conserving $L$ as well as $L'$, the scalar potential
of the model contains the soft symmetry breaking trilinear terms
\beq
\label{tril}
V_3(\phi_a,h,k) =\mu \left( \tilde{\phi}^{\dagger}_1\phi_2h^+ +
\phi_2^{\dagger} \tilde{\phi}_1 h^- \right) + \kappa \left( h^+h^+k^{--} +  h^-
h^-k^{++}\right) ,
\eeq
where $\tilde{\phi}_1=i\tau_2\phi^*_1$ is the doublet conjugate to $\phi_1$.
The $\mu$ term is characteristic of the Zee model \cite{zee} and  violates
$L$
by $\pm 2$ units, whereas the $\kappa$ term is characteristic of doubly
charged Higgs \cite{zee2, babu} and 
violates $L$ as well as $L'$ by $\pm 2$ units.

Neutrino masses are generated in the model through exchanges of the
physical charged Higgs bosons of the model. A linear combination of
$\phi_1^{\pm}$ and  $\phi_2^{\pm}$ is eaten up by $W^{\pm}$ while the
orthogonal combination ( $\equiv \phi^{\pm}$) mixes with the $h^{\pm}$ through
the $\m$ term.
The magnitude of this mixing is quite small, $
\approx \mu v/m_h^2$, for $v\equiv
\sqrt{\langle\phi_1^0\rangle^2 +
\langle\phi_2^0\rangle^2} << m_h^2/\mu$. We therefore continue to denote
the physical charged Higgs bosons as $h^{\pm}$ and $\phi^{\pm}$
 
At the one-loop level, the diagram in Fig. 1 contributes to the neutrino mass
matrix, which takes the form
\beq\label{numass}
M_{\n}^{(1)} = m_{\n} \left( \ba{ccc} 0 &\cos\theta_{\nu}&\sin\theta_{\nu}
\\
\cos\theta_{\nu}& 0 &0\\
 \sin\theta_{\nu}&0&0
\ea \right),
\eeq
which conserves $L'\equiv L_e-L_\m-L_\t$, while violating $L$ by two units. Here
\cite{smir2}
\beq\label{mnu}
m_{\nu} \approx \frac{\mu}{16\pi^2}\frac{\sqrt{f_{e\m}^2m_{\mu}^4 +
f_{e\t}^2m_{\t}^4}}{m_h^2}\log{\frac{m_h^2}{m_{\phi}^2}},
\eeq
and
\beq\label{tan}
\tan\theta_{\nu} \approx \frac{f_{e\t}m_{\t}^2}{f_{e\m}m_{\m}^2}.
\eeq
In deriving the above relations it is assumed 
that one of the Yukawa couplings ($g^{(1)}$) makes
the dominant contribution, and that $\langle \phi_1^0 \rangle \approx \langle 
\phi_2^0 \rangle$. Relaxing some of these simplifying assumptions would 
only give rise
to more free parameters, making it easier to fit the data.

The mass matrix $M_{\nu}$ has eigenvalues  $\pm m_{\nu}$, 0. At this stage
the
spectrum consists of a massless neutrino and a Dirac neutrino. For $m_{\nu}$ to
provide the scale of atmospheric neutrino oscillations, i.e., for $m_{\nu}
\approx 3\cdot 10^{-2}\, - \, 10^{-1}$ eV, we require, assuming $m_h \approx
10^4$ GeV and $\mu \approx \langle \phi_a^0 \rangle$,
\beq\label{femu}
\sqrt{f_{e\m}^2m_{\mu}^4 +f_{e\t}^2m_{\t}^4} \approx (0.4\, -\, 1.2)\cdot
10^{-4} \GeV^2.
\eeq
$M^{(1)}_{\nu}$ is diagonalized by a mixing matrix
\beq\label{unu}
U^{(1)}_{\nu} = \left( \ba{ccc}
1/\sqrt{2} & 1/\sqrt{2} & 0 \\
\cos\theta_{\nu}/\sqrt{2} & -\cos\theta_{\nu}/\sqrt{2} &-\sin\theta_{\nu}
\\
 \sin\theta_{\nu}/\sqrt{2} & -\sin\theta_{\nu}/\sqrt{2} &\cos\theta_{\nu} 
\ea
\right).
\eeq
This matrix displays the bimaximal structure \cite{bim} if
$\theta_\n=45^0$. This choice leads to the maximal amplitude for the 
$\nu_\m$ oscillations in conformity with the data on 
atmospheric neutrinos.
Eq. (\ref{tan}) then implies
\beq \label{tuned}
f_{e\t} m^2_{\t} \approx f_{e\m}m_{\m}^2. 
\eeq
Eqs. (\ref{femu}, \ref{tuned}) together require
\beq\label{fenos}
f_{e\m} \approx 4\cdot 10^{-1}, \, f_{e\t} \approx 10^{-3}.
\eeq

Having obtained the conditions for resolving the atmospheric neutrino anomaly
we can now examine how a two-loop contribution to the neutrino mass matrix can
provide a resolution of the solar neutrino problem.

The two-loop diagram contributing to the neutrino masses is shown in Fig. 2.
Assuming that the $k$ mass is dominant, 
the mass matrix at the two-loop level can
be estimated to be 
\beq\label{mnu2}
M_{\n}^{(2)} = \left( \ba{ccc} 
0 &m_{\n}\cos\theta_{\nu}&m_{\n}\sin\theta_{\nu} \\
m_{\n}\cos\theta_{\nu}& A f_{e\m}^2 &A f_{e\m} f_{e\t}\\
m_{\n} \sin\theta_{\nu}&A f_{e\m} f_{e\t} &A f_{e\t}^2
\ea \right),
\eeq
where
\beq \label{a1}
A \approx \frac{\kappa}{(16\pi^2)^2} h_{ee} \frac{m_e^2}{m_k^2},
\eeq
where we have assumed the logarithm factors to be of order unity. We see that
now the mass matrix $M_{\n}^{(2)}$ no longer respects the symmetry $L'$.
Consequently, the two eigenvalues ($\pm m_{\n}$) which were equal and opposite
at one-loop level, are now split by an amount $A f_{e\m}^2 \cos^2\theta_{\n}$.
This leads to a mass-squared difference
\beq \label{ds}
\D_S \approx 2 m_{\n}A f_{e\m}^2 \cos^2\theta_{\n}.
\eeq
For $\kappa\approx 10^2$ GeV, $m_k \approx 10^4$ GeV, and $h_{ee}
\approx 1$, this gives 
$\D_S \approx 5\cdot 10^{-11}$ \eV$^2$. This is in the range needed for the VO
solution.
It is of course obvious that the model does not permit a value of $\D_S$ in
the range for the MSW solutions. We will discuss later the possibility of
modifying the model to accommodate the LAMSW \cite{lamsw} solution.

We  thus see that there is a choice of parameters $f_{e\m}$, $f_{e\t}$,
$h_{ee}$, $\mu$, $\kappa$, $m_h$ and $m_k$ for which simultaneous solutions to
both atmospheric and solar neutrino anomalies can be found. Of these
parameters, the masses $m_h, m_k$ have to be somewhat large compared to the
weak scale, and the ratio $f_{e\t}/f_{e\m}$ has to be somewhat fine tuned.

We now look at constraints on the model coming from other experiments. These
have been studied in some detail in \cite{zee, smir2, babu, swar}. 
However, many of
these constraints are not relevant since many couplings vanish
automatically
because of the assumed $L'$ invariance. Thus, for example, since $k^{++}$
couples only to $e^+e^+$, unlike in  \cite{zee2,babu}, the only
constraints on its mass and coupling are from $(g-2)$ of the electron and from
the Bhabha scattering process $e^+e^-\rightarrow e^+e^-$. 
The former gives the limit \cite{babu}
\beq
\vert h_{ee} \vert ^2/ m_k^2 \ler 5 \cdot 10^{-2}\, {\rm GeV}^{-2},
\eeq
which is trivially satisfied for our choice of $h_{ee}$ and $m_k^2$. The limit
from Bhabha scattering is \cite{swar}
\beq
\vert h_{ee} \vert ^2/ m_k^2 \ler 9.7 \cdot 10^{-6}\, {\rm GeV}^{-2},
\eeq
which is also satisfied. 

So far as limits on the $h^+$ mass and couplings are concerned, these come from
data on $\mu$ decay together with $e-\m$ universality, and from $(g-2)$ of $e$
and $\mu$. Of these, the former one is more stringent \cite{babu}, viz., 
\beq
f_{e\m}^2/m_h^2 \ler 10^{-8}\, {\rm GeV}^{-2}.
\eeq
This constraint is satisfied for our choice of parameters.

It is straightforward to modify the above scheme so as to obtain the scale
 relevant
for the MSW solution. The only change needed is in the quantum number of
$k^{++}$ from $L'=-2$ to $L'=+2$. The $k^{++}$
Yukawa couplings in this case would be given by 
\beq
-{\cal L'}_Y  = 
\left[ h_{\m\m}\overline{\m^c_R}\m_L
+ h_{\m\t}\overline{\m^c_R}\t_L
+ h_{\t\m}\overline{\t^c_R}\m_L
+ h_{\t\t}\overline{\t^c_R}\t_L\right] k^{++}
 + {\rm H.c.}.
\eeq 
Now the degenerate neutrinos at 1-loop level are split by the diagram of Fig.2 
with
the dominant contribution coming from diagram with internal $\tau$ lepton. 
This has
the effect that under the simplifying assumptions of $h_{\m\t}=h_{\t\m}=0$, and
$h_{\m\m}
\approx h_{\t\t}$,   $A$ defined in eq. 
(\ref{a1}) would be replaced by 
\beq \label{a2}
A \approx \frac{\kappa}{(16\pi^2)^2} h_{\t\t} \frac{m_{\t}^2}{m_k^2}.
\eeq 
Consequently, $\Delta_S$ would be larger compared to its value in
eq.(\ref{ds}) due to $m_{\t}$ replacing  $m_e$ in the expression for $A$. 
Hence,
\beq
\D_S\approx 8\cdot 10^{-4} \; h_{\t\t}\;\; {\rm eV}^2.
\eeq
This can be in the right range \cite{lamsw} for the LAMSW solution for
$h_{\t\t}\sim
10^{-1}-10^{-2}$. 

The main obstacle in solving the solar neutrino problem in this case comes
from the mixing pattern in eq.(\ref{unu}). This implies  that the
mixing angle determining the vacuum survival probability of the
electron neutrino is exactly $45^\circ$ and radiative corrections to it
are too 
small to change it 
significantly.
It is well-known \cite{smir1,smir2} that the
MSW effect cannot take place inside the Sun in this case.
A possible way out is to 
make a different $L'$ assignment
for $\phi_2$, viz., $L'=2$. This has two consequences. 
Firstly, the charged lepton
mass matrix no longer remains \emt symmetric. Secondly, the \emt breaking in 
the charged lepton would induce similar breaking in the neutrino mass matrix.
Both these factor would  lead to deviation from the bimaximal structure 
in eq.(\ref{unu}). It is possible to choose parameters $g^2_{ij}$
in a way which would make the vacuum amplitude 
$4U_{e 1}^2 U_{e 2}^2$ for the oscillations of the solar neutrinos
significantly less than 1 
leading to a LAMSW solution to the solar neutrino problem.

To conclude, we have described an economical scenario which accommodates in a
natural way the smallness of the neutrino masses, as well as the hierarchy in
the scales $\D_S$ and $\D_A$ responsible for the understanding of the
solar and atmospheric neutrino data. In the model we propose, the vacuum
oscillation solution for the solar neutrino problem is natural. However, 
modified assignments of quantum numbers of scalars can give rise to a model
which would accommodate the large-angle MSW solution.

\newpage
\begin{figure}[t]
\epsfxsize 15 cm
\epsfysize 15 cm
\epsfbox[25 151 585 704]{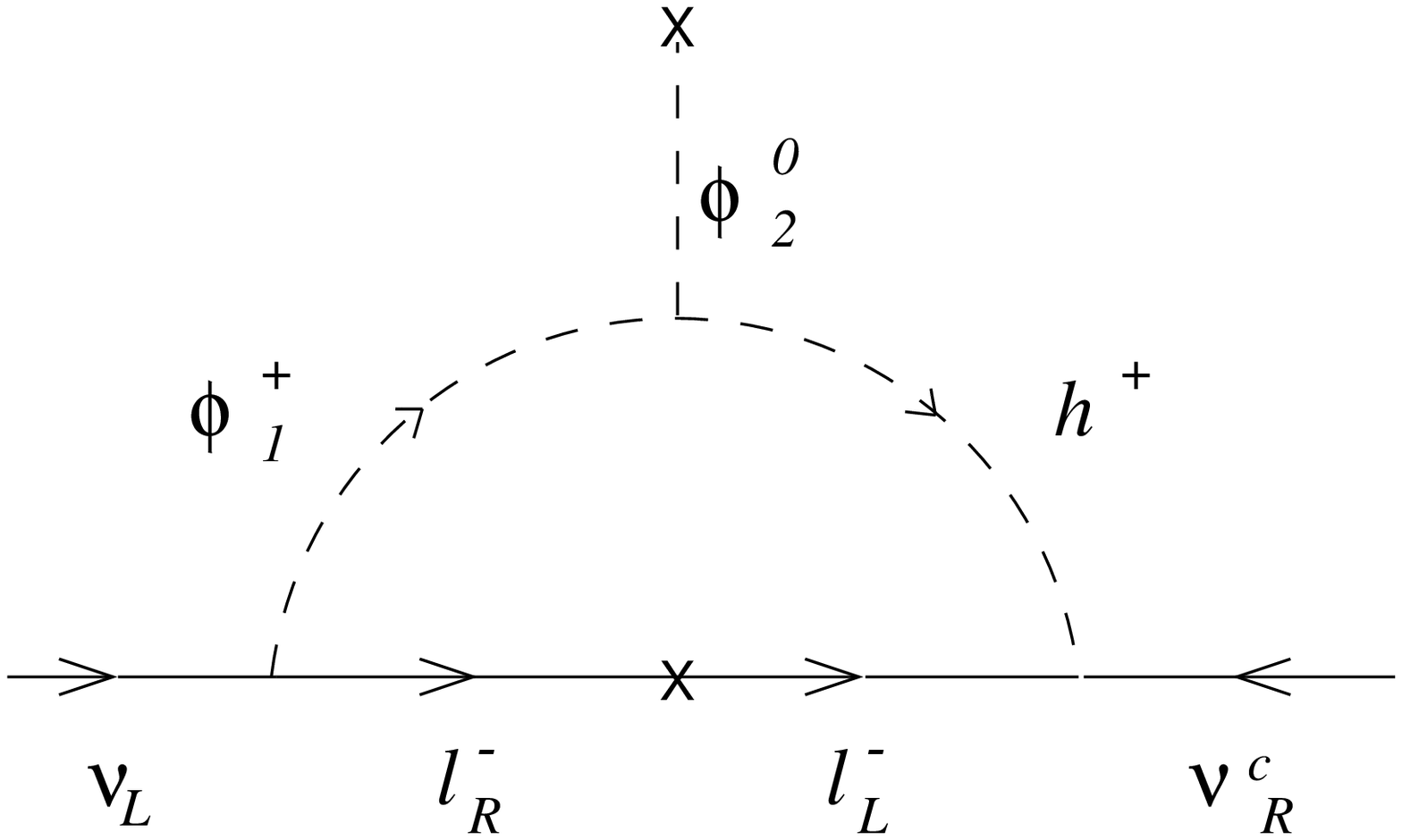}
\label{Fig. 1}
\caption {\sl One-loop diagram contributing to the 
neutrino mass matrix.}  
\end{figure}
\begin{figure}[t]
\epsfxsize 15 cm
\epsfysize 15 cm
\epsfbox[25 151 585 704]{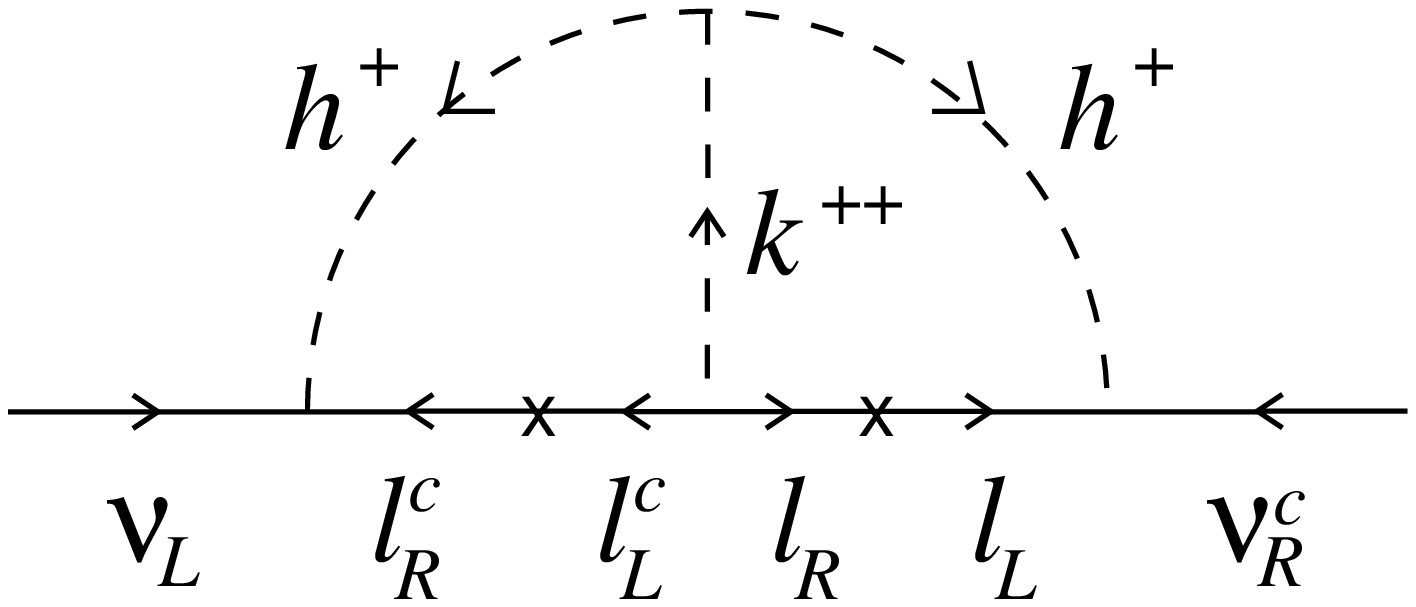}
\label{Fig. 2}
\caption {\sl Two-loop diagram contributing to the 
neutrino mass matrix.}  
\end{figure}
\end{document}